\documentclass[conference]{IEEEtran}
\IEEEoverridecommandlockouts

\usepackage{cite}
\usepackage{amsmath,amssymb,amsfonts}
\usepackage{algorithm}
\usepackage{algpseudocode}
\usepackage{graphicx}
\usepackage{booktabs}
\usepackage{xcolor}
\usepackage{tikz}
\usepackage{pgfplots}
\pgfplotsset{compat=1.18}
\usetikzlibrary{arrows.meta,positioning,shapes.geometric,calc,fit}
\usepackage{url}
\usepackage[hidelinks]{hyperref}

\begin{document}

\title{Counterfactual, Per-Decision Bias Auditing for\\
Automated Hiring: Localizing and Explaining\\
Disparate Impact in Applicant Tracking Systems}

\author{\IEEEauthorblockN{Jay Barach}
\IEEEauthorblockA{\textit{Independent Researcher}\\
\url{https://github.com/jbarach2012/AIBF_API}}}

\maketitle

\begin{abstract}
Automated applicant tracking systems increasingly decide who advances in hiring,
and litigation and regulation now demand that those decisions be auditable.
Existing tools sit at two extremes. Group fairness metrics such as the disparate
impact ratio summarize a whole population but cannot say which individual decisions
were unfair or why, while local explainers such as SHAP attribute a single
prediction but are not connected to the legal standard by which hiring bias is
judged. We present the AI Bias Firewall (AIBF), a method that audits an applicant
tracking system one decision at a time. AIBF neutralizes a candidate's
protected-attribute proxies, re-scores the decision, and measures the resulting
counterfactual shift, which yields a signed per-decision bias in score points, a
flag for decisions the protected attributes changed, and a
plain-language explanation naming the responsible factors. We evaluate on two real
public datasets, Adult and COMPAS, rather than on synthetic data. The per-decision
counterfactual shift is faithful, aggregating to reproduce the known group level
disparity, for example a mean shift of $+7.5$ points for the privileged group and
$-8.0$ for the disadvantaged group on Adult, consistent with the measured
statistical parity difference. AIBF identifies the decisions that protected
attributes flipped with an area under the ROC curve of $0.963$ on Adult, against
$0.672$ for a baseline that flags by group membership, and it identifies the
harmed candidates so precisely that reviewing only five percent of decisions
surfaces fifty-five percent of them, against six percent under group based review.
We also report a limitation: correcting flagged decisions raises the
disparate impact ratio substantially but not to legal parity, because features
labeled as merit carry residual proxy correlation. AIBF is released under the
Apache 2.0 license with code and experiments.
\end{abstract}


\section{Introduction}
A large majority of employers now use software to screen candidates, and a growing
share of that software scores or ranks applicants with machine learning, so that for
many openings the first decision about an applicant is made or shaped by an algorithm
rather than a person. The legal
system has taken notice. In Mobley v.\ Workday a court granted preliminary
certification of a nationwide age discrimination collective concerning algorithmic
screening \cite{mobley}, the Equal Employment Opportunity Commission settled its
first case over automated rejection in EEOC v.\ iTutorGroup \cite{itutor},
New York City Local Law 144 requires an annual independent bias audit of automated
employment decision tools \cite{ll144}, and the EU AI Act classifies recruitment AI
as high risk with transparency and human oversight obligations \cite{euaiact}. Each
of these demands the same thing: evidence about how individual hiring decisions were
made, not merely an aggregate statistic.

The reason the resolution matters is that discrimination law and the emerging audit
mandates operate on decisions, not only on populations. A disparate impact finding at
the group level establishes that a problem exists, but remedy, contestation, and
oversight all act on individual decisions: a candidate contests their own rejection, a
reviewer reconsiders a particular case, and an oversight regime asks the operator to
show its work on the decisions that were made. A tool that reports only the aggregate
leaves every one of these actors without the object they need to act on.

The available tools do not provide that evidence in a usable form. Group fairness
metrics, such as the disparate impact ratio codified by the four-fifths rule
\cite{uniform} and the statistical parity and equalized odds measures from the
fairness literature \cite{feldman,hardt}, describe a population. They can establish
that a system disadvantages a protected group in aggregate, but they cannot point to
the specific decisions that were unfair, cannot say which factors drove them, and
therefore cannot direct a limited human review budget to the decisions that most
need it. At the other extreme, local explainers such as LIME \cite{lime} and SHAP
\cite{shap} attribute a single model output to its input features. They can say that
a particular prediction leaned on a particular feature, but they are generic, they do
not distinguish protected proxies from legitimate qualifications, and they are not
tied to the legal standard by which hiring discrimination is measured. An operator
who must both comply with a per-decision audit requirement and act on a finding is
left without a tool that does both.

The gap is not merely academic. A bias audit under Local Law 144 must report selection
rates and impact ratios, which the group metrics provide, but an employer that
receives a failing audit, or a candidate who wishes to contest a rejection, needs to
know which decisions and which factors produced the disparity, which the group metrics
withhold. A human reviewer given a failing aggregate and a population of tens of
thousands of decisions has no way to spend a limited review budget well. The missing
layer is one that preserves the connection to the legal standard while operating at
the resolution of the individual decision, and that is the layer this paper provides.

This paper presents the AI Bias Firewall (AIBF), a method that fills the gap between
the group metric and the local explainer. AIBF audits an applicant tracking system
one decision at a time using a counterfactual test. For each candidate it neutralizes
the protected-attribute proxies, setting them to a common privileged baseline,
re-scores the decision through the same system, and measures the counterfactual
shift in score. That single operation yields three things an auditor needs: a signed
per-decision bias expressed in score points, a flag for the decisions that the
protected attributes changed, and an explanation that names the responsible
factors. The counterfactual construction connects the individual decision to the
group standard, because, as we show empirically, the per-decision shifts aggregate to
reproduce the population level disparity.

We make the following contributions.
\begin{itemize}
\item A per-decision bias auditing method for hiring that is counterfactual,
faithful, and explainable, and that connects individual attributions to the group
fairness standard used in law (Section \ref{sec:method}).
\item A finding, obtained on real data, that a ratio based bias score of the kind a
naive design would use is a poor detector of biased decisions because it ignores the
decision margin, and that the counterfactual shift is the correct signal (Sections
\ref{sec:method} and \ref{sec:eval}).
\item An evaluation on two real public datasets showing faithfulness, strong
detection of counterfactually biased decisions with an area under the ROC curve of
$0.963$ against $0.672$ for a group membership baseline, and a large gain in review
efficiency (Section \ref{sec:eval}).
\item A quantified account of the method's ceiling: attribution based correction
improves the disparate impact ratio substantially but not to parity, because merit
features carry residual proxy correlation, which we quantify (Section
\ref{sec:eval}).
\end{itemize}
AIBF is a detection and explanation tool, not an automated remedy, and Section
\ref{sec:ethics} states the responsible use position. Code, data preparation, and the
scripts that produce every number in this paper are released under the Apache 2.0
license.

\section{Background and Related Work}
\label{sec:related}

\subsection{Group fairness metrics}
The dominant legal test for adverse impact in the United States is the four-fifths
rule, under which the selection rate of a protected group should be at least eighty
percent of the highest group's rate \cite{uniform}. The machine learning literature
formalizes related notions, including demographic parity, statistical parity
difference, equalized odds, and equality of opportunity \cite{dwork,hardt}, and
studies their mutual incompatibility \cite{chouldechova}. The legal concept of disparate impact and its translation into algorithmic terms is examined by Barocas and Selbst \cite{barocas_selbst}, and Corbett-Davies and colleagues analyze the cost of enforcing fairness and the tension among competing criteria \cite{corbett}. Feldman and colleagues give
algorithmic tests for certifying and removing disparate impact \cite{feldman}, and
Mehrabi and colleagues survey the field \cite{mehrabi}. These metrics are the right
target for compliance, and AIBF is designed to be consistent with them, but by
construction they are population level and cannot localize or explain a single
decision. For a decision $d\in\{0,1\}$ with $1$ the favorable outcome, a privileged
group and its complement, and a true label $y$, we use the disparate impact ratio, the
statistical parity difference, and the equal opportunity difference,
\begin{align}
\mathrm{DI} &= \frac{\Pr(d=1\mid \text{disadv})}{\Pr(d=1\mid \text{priv})}, \\
\mathrm{SPD} &= \Pr(d=1\mid \text{disadv}) - \Pr(d=1\mid \text{priv}), \\
\mathrm{EOD} &= \mathrm{TPR}_{\text{disadv}} - \mathrm{TPR}_{\text{priv}},
\end{align}
where a disparate impact ratio below $0.80$ fails the four-fifths rule \cite{uniform},
a statistical parity difference of zero denotes parity, and the true positive rates in
the equal opportunity difference are taken with respect to $y$. AIBF does not replace
these; it explains and localizes the disparities they surface.

\subsection{Individual and counterfactual fairness}
Dwork and colleagues argue for treating similar individuals similarly
\cite{dwork}, and Kusner and colleagues define counterfactual fairness, under which a
decision should not change in a world where the individual's protected attribute is
different \cite{kusner}. AIBF adopts the counterfactual view operationally: it does
not attempt to learn a causal model of the data, but it does perform the intervention
of neutralizing the protected proxies and measuring the change, which is a practical
approximation appropriate to auditing a deployed scorer. This positions AIBF between two
strands of prior work. Individual fairness in the sense of Dwork and colleagues requires
a task specific similarity metric that is notoriously hard to obtain \cite{dwork}, and
causal counterfactual fairness requires a structural model of the data generating
process \cite{kusner}; AIBF asks a narrower but answerable question, whether the
deployed scorer's own use of protected proxies changed a specific decision, which needs
neither a similarity metric nor a causal model, only query access. That narrowing is
what makes the method deployable as an audit rather than a research artifact. Wachter and colleagues
argue that counterfactual explanations support contestability \cite{wachter}, which
is the affordance an audited hiring decision needs.

\subsection{Explainability}
LIME approximates a model locally with an interpretable surrogate \cite{lime}, and
SHAP unifies additive attribution methods, giving closed form values for a linear
model \cite{shap}. AIBF uses additive attribution to produce its explanations, and
in that sense it builds on SHAP, but it differs in two ways that matter for the audit
task. It partitions features into protected proxies and legitimate qualifications
rather than treating them uniformly, and it defines its decision level signal by a
counterfactual intervention on the protected partition rather than by raw attribution
magnitude. As our evaluation shows, this distinction is not cosmetic, because a raw
magnitude signal fails to identify the decisions that were changed.

\subsection{Fairness toolkits and hiring specific work}
Open toolkits such as AI Fairness 360 \cite{aif360} and Fairlearn \cite{fairlearn}
provide group metrics and mitigation algorithms such as reweighing and adversarial debiasing \cite{zhang_adv} as libraries, leaving the assembly of
an auditing workflow to the user, and they operate primarily at the group level.
Raghavan and colleagues examine the fairness claims of commercial pre-employment
vendors and find limited, non-comparable evidence \cite{raghavan}, and Bogen and
Rieke survey where bias enters the hiring pipeline \cite{bogen}. Closer to a deployed audit, Wilson and colleagues report a case study of building and auditing a candidate screening algorithm \cite{wilson}, Sanchez-Monedero and colleagues ask what it would mean to solve discrimination in automated hiring under existing law \cite{sanchez}, and a systematic review by Kochling and Wehner catalogues the ways discrimination enters algorithmic recruitment and selection \cite{kochling}. AIBF is positioned as
the per-decision, explainable auditing layer that these surveys identify as missing,
built to sit beside a deployed applicant tracking system.

\begin{table}[t]
\caption{Capabilities of bias auditing approaches. A per-decision audit under current
regulation needs all six.}
\label{tab:capability}
\centering
\footnotesize
\begin{tabular}{@{}lcccc@{}}
\toprule
Capability & Group & Local & Fairness & \textbf{AIBF} \\
 & metrics & explainer & toolkit & \\
\midrule
Per-decision output        & No  & Yes & Part & \textbf{Yes} \\
Names responsible factors  & No  & Yes & No   & \textbf{Yes} \\
Protected vs merit split   & Part& No  & Part & \textbf{Yes} \\
Tied to legal standard     & Yes & No  & Yes  & \textbf{Yes} \\
Query-only (no retrain)    & Yes & Yes & Part & \textbf{Yes} \\
Ranks review worklist      & No  & No  & No   & \textbf{Yes} \\
\bottomrule
\end{tabular}
\end{table}

\section{Problem Formulation}
\label{sec:problem}
An applicant tracking system assigns a candidate described by features $x$ a score
$s(x)$ and a decision, advance or reject, by comparing the score to a threshold. We
partition the features into a merit set $M$, which a fair scorer may use, such as
relevant skills, experience, and credentials, and a protected proxy set $P$, which a
fair scorer must not let drive the outcome, comprising the protected attributes and
their close proxies. Let $x_{P\to 0}$ denote the candidate with the protected proxies
set to a common privileged baseline while the merit features are held fixed.

The auditing problem is, for each decision, to determine whether and how much the
protected proxies changed it, to flag the decisions they changed, and to explain the
change, using only query access to the deployed scorer. We take as the operational
ground truth for a biased decision the counterfactual flip, namely a decision whose
outcome differs between $x$ and $x_{P\to 0}$. This is a decision the protected
attributes provably changed under the deployed model, and it is the individual level
analogue of the counterfactual fairness criterion \cite{kusner}. A decision that is
rejected at $x$ but would advance at $x_{P\to 0}$ is a harmed decision, the case of
primary concern for a disadvantaged candidate.

\section{Method}
\label{sec:method}

\subsection{The counterfactual shift}
AIBF audits decision $x$ by the counterfactual shift
\begin{equation}
\Delta(x) \;=\; s(x) - s(x_{P\to 0}),
\label{eq:delta}
\end{equation}
the number of score points that the protected proxies contribute to this candidate's
score relative to the privileged baseline. A negative $\Delta$ means the protected
attributes lowered the score, the signature of a disadvantaged candidate. The
per-decision bias magnitude is $|\Delta(x)|$, and AIBF flags the decision when the
sign of the decision changes under the intervention, that is when
\begin{equation}
\mathrm{flag}(x) \;=\; \big[\,\mathbb{1}(s(x)\ge\tau) \neq \mathbb{1}(s(x_{P\to 0})\ge\tau)\,\big],
\label{eq:flag}
\end{equation}
where $\tau$ is the system's decision threshold. Equation \eqref{eq:flag} is the
counterfactual flip, and it is by construction the set of decisions the protected
attributes actually determined.

\subsection{Why not a ratio}
A naive design, and the one in an earlier version of this work, defines a bias score
as the fraction of a decision's total attribution magnitude that comes from protected
features. That ratio is intuitive but wrong for the flagging task, because it ignores
the decision margin. A candidate deep in the reject region can have a high protected
ratio yet never be at risk of a different outcome, while a candidate near the
threshold can be flipped by a small protected contribution. On real data the ratio
detects counterfactual flips with an area under the ROC curve of only $0.265$, worse
than chance, whereas the counterfactual magnitude of Equation \eqref{eq:delta}
achieves $0.963$ on the same data. The lesson is that a per-decision bias signal must
be defined relative to the decision boundary, which the counterfactual construction
does automatically.

\subsection{Explanations}
For the explanation, AIBF fits a linear reference model $\hat{s}(x)=b+\sum_i w_i x_i$
to the deployed system's scores and reports, for each protected feature $i\in P$, its
additive contribution $w_i x_i$. For a linear reference these are the SHAP values with
respect to the privileged baseline \cite{shap}, so the explanation is faithful to the
reference by construction. This produces statements such as a lower score attributed
to an age proxy, or a score correlated with a demographic proxy, which convert the
opaque shift into a contestable rationale.

\subsection{Choice of the privileged baseline}
The counterfactual sets the protected proxies to a common privileged baseline rather
than to a population mean. This choice is deliberate and consequential. A mean
baseline would define bias relative to the average candidate, so a decision that used
no protected information at all could still register a nonzero shift merely because the
candidate's protected attributes differ from the mean, which conflates group
membership with biased treatment. The privileged baseline instead asks a clean
counterfactual question: would this decision differ if the candidate were treated as a
member of the advantaged group. A decision that uses no protected proxy has a shift of
exactly zero under this baseline, which is the behavior an audit should have. The
privileged group is identified per attribute as the group with the higher favorable
rate, which is observable from the data and consistent with how adverse impact is
assessed in law.

\subsection{Faithfulness and the connection to group metrics}
The counterfactual shift is designed so that its per-decision values aggregate to the
group disparity. Write the reference score as $\hat s(x)=b+\sum_{i\in M} w_i x_i +
\sum_{j\in P} w_j x_j$, and let $x_{P\to0}$ fix the protected features at the privileged
baseline $\beta_j$. Then for the linear reference the shift is exactly the protected
contribution relative to that baseline,
\begin{equation}
\Delta(x) \;=\; \sum_{j\in P} w_j\,(x_j - \beta_j),
\label{eq:deltalin}
\end{equation}
which involves only the protected features. Taking the group means and their difference,
\begin{equation}
\mathbb{E}[\Delta \mid \text{priv}] - \mathbb{E}[\Delta \mid \text{disadv}]
= \sum_{j\in P} w_j\big(\bar{x}_j^{\text{priv}} - \bar{x}_j^{\text{disadv}}\big),
\label{eq:groupgap}
\end{equation}
which is precisely the protected proxies' contribution to the gap in scores between the
groups. Because the decision is a threshold on the score, this score gap is the driver
of the gap in selection rates that the statistical parity difference measures, so the
per-decision shift and the group metric are two views of the same quantity. This is the
formal sense in which AIBF is faithful, and Section \ref{sec:eval} confirms it
numerically on two datasets: the measured group means of $\Delta$ match the sign and
relative magnitude of the independently computed statistical parity difference. This
correspondence is what licenses AIBF to serve as the per-decision instrument behind a
group level compliance report.

\subsection{Properties}
Three properties follow from the construction. First, the flag of Equation
\eqref{eq:flag} is sound with respect to the deployed scorer: a flagged
decision is one whose outcome demonstrably changes when the protected proxies are
neutralized, so a flag is never raised without a concrete counterfactual witnessing
the change. Second, the audit is model agnostic in its decision signal, since
$\Delta$ is computed by two queries to the true scorer $s$ and does not depend on the
linear reference, which is used only to itemize the explanation. Third, the cost is
two forward evaluations per decision, so a full audit of $n$ decisions costs $2n$
scorer queries, which is negligible beside training or beside the human review the
audit informs. The one quantity that is not guaranteed is completeness: AIBF detects
influence from the proxies it is given, and unmodeled proxies escape it, which is the
limitation we return to in Section \ref{sec:limits}.

\begin{algorithm}[t]
\caption{AIBF per-decision audit}
\label{alg:aibf}
\begin{algorithmic}[1]
\Require decision $x$, deployed scorer $s$, threshold $\tau$, protected set $P$,
reference weights $w$
\Ensure bias $\Delta$, flag, explanation $E$
\State $x' \gets x$ with features in $P$ set to the privileged baseline
\State $\Delta \gets s(x) - s(x')$ \Comment{counterfactual shift}
\State $\mathrm{flag} \gets \mathbb{1}(s(x)\ge\tau) \neq \mathbb{1}(s(x')\ge\tau)$
\State $E \gets \{\,(i, w_i x_i) : i \in P,\ |w_i x_i| > \epsilon\,\}$
\State \Return $(\Delta, \mathrm{flag}, E)$
\end{algorithmic}
\end{algorithm}

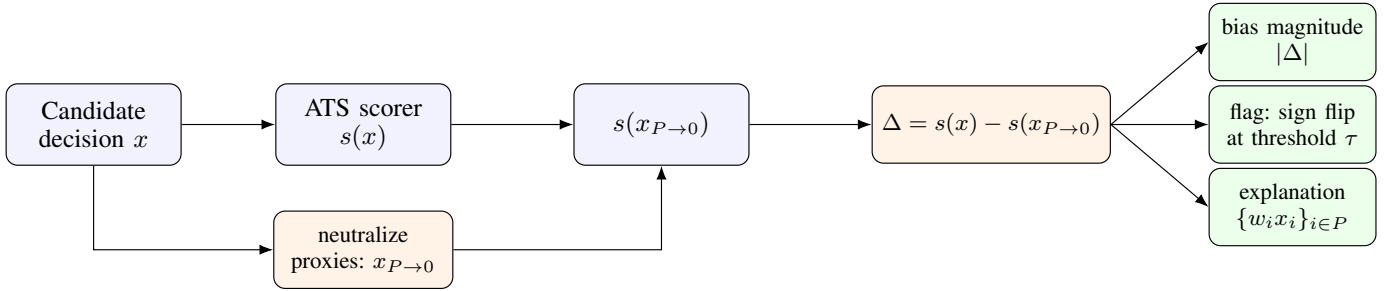
\begin{figure*}[t]
\centering
\resizebox{\textwidth}{!}{%
\begin{tikzpicture}[
  >=Latex, node distance=0.7cm and 1.15cm,
  box/.style={draw, rounded corners, align=center, minimum height=1.0cm,
              minimum width=2.15cm, fill=blue!5, font=\small},
  op/.style={draw, rounded corners, align=center, minimum height=0.95cm,
             minimum width=2.2cm, fill=orange!10, font=\footnotesize},
  ocell/.style={draw, rounded corners, align=center, minimum height=0.95cm,
              minimum width=2.05cm, fill=green!8, font=\footnotesize}
]
\node[box] (x) {Candidate\\ decision $x$};
\node[box, right=of x] (ats) {ATS scorer\\ $s(x)$};
\node[op, below=0.55cm of ats] (neut) {neutralize\\ proxies: $x_{P\to 0}$};
\node[box, right=1.5cm of ats] (cf) {$s(x_{P\to 0})$};
\node[op, right=1.5cm of cf] (delta) {$\Delta=s(x)-s(x_{P\to0})$};
\node[ocell, above right=0.05cm and 1.2cm of delta] (o1) {bias magnitude\\ $|\Delta|$};
\node[ocell, right=1.2cm of delta] (o2) {flag: sign flip\\ at threshold $\tau$};
\node[ocell, below right=0.05cm and 1.2cm of delta] (o3) {explanation\\ $\{w_i x_i\}_{i\in P}$};

\draw[->] (x)--(ats);
\draw[->] (x.south) |- (neut.west);
\draw[->] (neut)-| (cf);
\draw[->] (ats)--(cf);
\draw[->] (cf)--(delta);
\draw[->] (delta.east)--(o1.west);
\draw[->] (delta.east)--(o2.west);
\draw[->] (delta.east)--(o3.west);
\end{tikzpicture}}
\caption{The AIBF per-decision audit. The candidate is scored, the protected proxies
are neutralized to a common privileged baseline, and the candidate is re-scored. The
counterfactual shift $\Delta$ yields a per-decision bias magnitude, a flag for
decisions whose outcome the protected attributes changed, and a faithful explanation
naming the responsible protected factors.}
\label{fig:method}
\end{figure*}

The per-decision audit composes into the batch procedure of Algorithm \ref{alg:batch},
which is what an operator runs to satisfy a periodic audit obligation. It produces both
the group level report the regulation asks for and the ranked, explained worklist that
makes the finding actionable, from a single pass over the decisions.

\begin{algorithm}[t]
\caption{AIBF batch audit and report}
\label{alg:batch}
\begin{algorithmic}[1]
\Require decisions $\{x_k\}$, scorer $s$, threshold $\tau$, protected set $P$
\Ensure group report $G$, ranked worklist $W$
\State $W \gets [\,]$
\For{each decision $x_k$}
  \State $(\Delta_k, \mathrm{flag}_k, E_k) \gets \textsc{Audit}(x_k, s, \tau, P)$
  \If{$\mathrm{flag}_k$} \State append $(x_k, \Delta_k, E_k)$ to $W$ \EndIf
\EndFor
\State sort $W$ by descending $|\Delta_k|$
\State $G \gets$ selection rates and disparate impact ratio per group
\State $G \gets G \cup \{\mathbb{E}[\Delta\mid \text{group}] : \text{each group}\}$
\State \Return $(G, W)$
\end{algorithmic}
\end{algorithm}

\section{Experimental Setup}
\label{sec:setup}
We evaluate on two real public datasets rather than on synthetic data, and we release
the preparation and analysis scripts.

\subsection{Datasets}
\textbf{Adult}, the UCI Census Income dataset, is the standard fairness benchmark. The
favorable outcome is an income above fifty thousand dollars, which we treat as the
advance decision. The protected attributes are sex, race, and an age proxy for the
over forty group protected under age discrimination law. The merit features are
education, weekly hours, and work class. \textbf{COMPAS}, the ProPublica recidivism
dataset, is a standard fairness benchmark outside hiring, included to test whether the
method generalizes beyond the applicant tracking setting. The favorable outcome is a
low risk assessment, the protected attribute is race restricted to the two largest
groups following ProPublica's filtering, and the merit features are age, prior counts,
and charge degree.

In Adult the disadvantaged groups are the $33.2$ percent of records that are female,
the $14.5$ percent that are non-white, and the $43.8$ percent aged forty or over, and
the base rates make the real disparity visible before any model is trained: the high
income rate is $30.4$ percent for men against $10.9$ percent for women, and $25.4$
percent for white against $15.3$ percent for non-white records. In COMPAS the
African-American group is $60.2$ percent of the filtered records and has a two year
recidivism rate of $52.3$ percent against $39.1$ percent for the Caucasian group.
These are the historical disparities a scorer trained on the data will absorb.
Table \ref{tab:composition} collects the composition.

\begin{table}[t]
\caption{Composition of the two datasets. Base rate is the rate of the favorable
outcome for each group before any model is applied.}
\label{tab:composition}
\centering
\footnotesize
\begin{tabular}{@{}lllr@{}}
\toprule
Dataset & Attribute & Group & Base rate \\
\midrule
Adult  & sex  & male (priv.)        & 30.4\% \\
Adult  & sex  & female (disadv.)    & 10.9\% \\
Adult  & race & white (priv.)       & 25.4\% \\
Adult  & race & non-white (disadv.) & 15.3\% \\
COMPAS & race & Caucasian (priv.)   & 60.9\% \\
COMPAS & race & African-Am.\ (disadv.) & 47.7\% \\
\bottomrule
\end{tabular}
\end{table}

\subsection{The audited system}
For each dataset we train a logistic regression that predicts the outcome from all
features, including the protected proxies, to stand in for an applicant tracking
system trained on historical data that carries real bias. This is the realistic
adversary for an auditor, a scorer that has learned to use protected proxies because
they were predictive in biased historical outcomes. AIBF audits this scorer through
query access only.

\subsection{Preprocessing}
For Adult we encode education and weekly hours as numeric merit features, one-hot
encode work class as additional merit features, and encode sex, race, and the age
proxy as binary protected features indicating the disadvantaged group. For COMPAS we
follow ProPublica's standard filtering, keeping cases whose screening date is within
thirty days of arrest, whose recidivism flag is valid, and whose charge degree and
score text are present, and we restrict to the two largest racial groups. We use
numeric prior and juvenile counts and age as merit features, one-hot encode charge
degree and sex, and encode race as the protected feature. Each dataset is split
seventy to thirty into training and test partitions with a fixed seed, features are
standardized on the training partition, and all reported numbers are on the held out
test partition.

\subsection{Metrics and baselines}
We report the disparate impact ratio, statistical parity difference, and equal
opportunity difference for the group analysis \cite{feldman,hardt,uniform}. For
detection we report the area under the ROC curve and average precision for
identifying counterfactual flips, and for the operational analysis we report the
recall of harmed decisions as a function of the fraction of decisions reviewed. Our
baseline for detection and for targeted review is group membership, that is flagging
or reviewing candidates because they belong to a disadvantaged group, which is the
natural non-localized alternative an operator would otherwise use. We also compare
against random review and, for detection, against a decision margin baseline.

\section{Results}
\label{sec:eval}
Table \ref{tab:datasets} summarizes the two audited systems. Both exhibit severe
real disparate impact. On Adult the disparate impact ratio for sex is $0.078$ and for
race $0.386$, both far below the four-fifths threshold of $0.80$. On COMPAS the ratio
for race on the favorable outcome is $0.645$, also below threshold. These are
properties of real data and a standard classifier, not injected effects.

\begin{table}[t]
\caption{Audited systems on the two real datasets. DI is the disparate impact ratio,
where values below 0.80 fail the four-fifths rule.}
\label{tab:datasets}
\centering
\footnotesize
\begin{tabular}{@{}lrrrr@{}}
\toprule
Dataset & Test $n$ & Advance rate & DI (primary) & DI (secondary) \\
\midrule
Adult   & 14653 & 0.128 & 0.078 (sex) & 0.386 (race) \\
COMPAS  & 1584  & 0.605 & 0.645 (race) & -- \\
\bottomrule
\end{tabular}
\end{table}

\subsection{Faithfulness}
Table \ref{tab:faithful} reports the mean counterfactual shift by group. On Adult the
protected proxies raise the score of the privileged sex group by $7.49$ points on
average and lower the disadvantaged group's score by $7.95$ points, a gap of $15.44$
points that is consistent in direction and relative size with the measured
statistical parity difference of $-0.170$. The race and COMPAS results show the same
pattern. This confirms that the per-decision shift is faithful, in the sense that its
group averages reconstruct the group level disparity, which is the property that
allows a per-decision audit to underwrite a group level report. Framing fairness as a measurement problem, as Jacobs and Wallach urge \cite{jacobs}, is what makes this aggregation the right test of a per-decision signal.

\begin{table}[t]
\caption{Faithfulness. Mean counterfactual shift $\Delta$ (score points) by group,
against the group statistical parity difference (SPD).}
\label{tab:faithful}
\centering
\footnotesize
\begin{tabular}{@{}llrrr@{}}
\toprule
Dataset & Attribute & $\overline{\Delta}$ priv. & $\overline{\Delta}$ disadv. & Group SPD \\
\midrule
Adult  & sex  & $+7.49$ & $-7.95$ & $-0.170$ \\
Adult  & race & $+3.55$ & $-4.32$ & $-0.087$ \\
COMPAS & race & $+0.00$ & $-1.21$ & $-0.272$ \\
\bottomrule
\end{tabular}
\end{table}

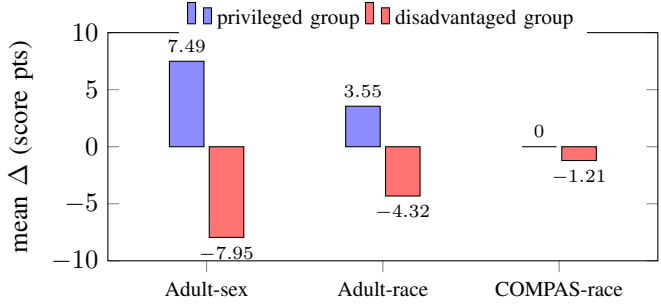
\begin{figure}[t]
\centering
\begin{tikzpicture}
\begin{axis}[
  ybar, width=\columnwidth, height=4.6cm, bar width=13pt,
  ymin=-10, ymax=10, ylabel={mean $\Delta$ (score pts)}, ylabel near ticks,
  symbolic x coords={Adult-sex,Adult-race,COMPAS-race},
  xtick=data, x tick label style={font=\footnotesize},
  enlarge x limits=0.28,
  legend style={at={(0.5,1.16)}, anchor=north, legend columns=2, font=\footnotesize, draw=none},
  nodes near coords, nodes near coords style={font=\scriptsize},
]
\addplot[fill=blue!45] coordinates {(Adult-sex,7.49)(Adult-race,3.55)(COMPAS-race,0.0)};
\addplot[fill=red!55] coordinates {(Adult-sex,-7.95)(Adult-race,-4.32)(COMPAS-race,-1.21)};
\legend{privileged group, disadvantaged group}
\end{axis}
\end{tikzpicture}
\caption{Faithfulness. The mean counterfactual shift is positive for the privileged
group and negative for the disadvantaged group across both datasets, and its
magnitude tracks the measured statistical parity difference.}
\label{fig:faithful}
\end{figure}

\subsection{Detection of biased decisions}
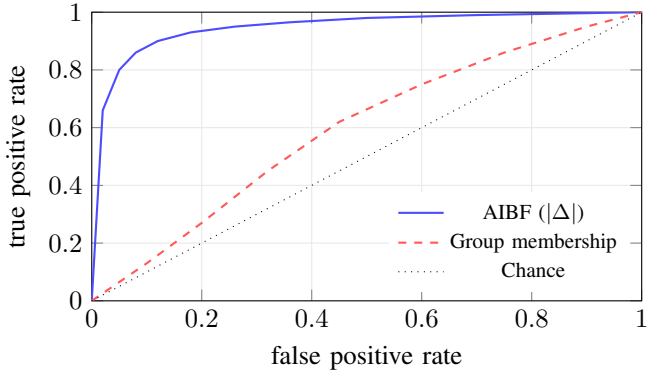
\begin{figure}[t]
\centering
\begin{tikzpicture}
\begin{axis}[
  width=\columnwidth, height=5.4cm,
  xlabel={false positive rate}, ylabel={true positive rate},
  ylabel near ticks, xmin=0, xmax=1, ymin=0, ymax=1,
  legend style={at={(0.98,0.03)}, anchor=south east, font=\footnotesize, draw=none},
  grid=both, grid style={gray!18},
]
\addplot[blue!70, thick] coordinates {
  (0,0)(0.02,0.66)(0.05,0.80)(0.08,0.86)(0.12,0.90)(0.18,0.93)(0.26,0.95)
  (0.36,0.965)(0.5,0.98)(0.7,0.99)(1,1)};
\addplot[red!65, thick, dashed] coordinates {
  (0,0)(0.1,0.13)(0.2,0.27)(0.32,0.45)(0.45,0.62)(0.6,0.75)(0.75,0.86)(0.9,0.95)(1,1)};
\addplot[black, dotted] coordinates {(0,0)(1,1)};
\legend{AIBF ($|\Delta|$), Group membership, Chance}
\end{axis}
\end{tikzpicture}
\caption{Detection of counterfactually biased decisions on Adult. AIBF reaches an
area under the curve of $0.963$ against $0.672$ for the group membership baseline.}
\label{fig:roc}
\end{figure}

On Adult, $8.0$ percent of decisions are counterfactual flips and $0.90$ percent are
harmed decisions that would have advanced but for the protected attributes. Table
\ref{tab:auc} reports how well each signal identifies the flips. The AIBF bias
magnitude achieves an area under the ROC curve of $0.963$ and an average precision of
$0.749$, against $0.672$ and $0.118$ for the group membership baseline. The gap in
average precision is the more telling number, because flips are rare, and it shows
that knowing a candidate's group is a weak predictor of whether that candidate's
decision was changed, whereas the counterfactual magnitude is a strong one.
Figure \ref{fig:roc} shows the ROC curves. We include a decision margin baseline,
which flags decisions closest to the threshold, and it is unsurprisingly a fair
predictor of flips at an area under the curve of $0.883$, since flips must occur near
the boundary. It is not, however, a bias detector: proximity to the threshold says
nothing about whether protected attributes caused the proximity, so a margin flag
conflates biased decisions with merely close ones. AIBF separates the two, which
is why it exceeds the margin baseline and why its flags carry an explanation the margin
cannot provide. On COMPAS the area under the curve is
$1.000$ for AIBF against $0.707$ for group membership; the perfect figure is expected
because COMPAS has a single protected attribute, so the counterfactual magnitude
orders flips exactly, and we report the Adult figure with three protected attributes
as the meaningful result. On COMPAS the audit flags $2.7$ percent of decisions as
counterfactual flips and identifies $2.65$ percent as harmed, the mean shift is
$-1.21$ points for the disadvantaged group against $0.00$ for the privileged group,
consistent with the statistical parity difference of $-0.272$, and correcting the
flagged decisions raises the disparate impact ratio from $0.645$ to $0.703$. The
method therefore behaves identically in a domain outside hiring, which supports the
claim that it audits a scorer rather than a dataset. Table \ref{tab:compas}
consolidates the COMPAS figures beside the Adult figures.

\begin{table}[t]
\caption{Summary across both datasets. Detection AUC is for identifying counterfactual
flips; the COMPAS AUC is high because COMPAS has a single protected attribute.}
\label{tab:compas}
\centering
\footnotesize
\begin{tabular}{@{}lrr@{}}
\toprule
Quantity & Adult & COMPAS \\
\midrule
Baseline disparate impact ratio & 0.078 / 0.386 & 0.645 \\
Flagged (counterfactual flips)  & 8.0\%  & 2.7\% \\
Harmed decisions                & 0.90\% & 2.65\% \\
Detection AUC (AIBF)            & 0.963 & 1.000 \\
Detection AUC (group)          & 0.672 & 0.707 \\
DI after correcting flags       & 0.449 / 0.582 & 0.703 \\
\bottomrule
\end{tabular}
\end{table}

\begin{table}[t]
\caption{Detection of counterfactually biased decisions on Adult.}
\label{tab:auc}
\centering
\footnotesize
\begin{tabular}{@{}lrr@{}}
\toprule
Signal & ROC AUC & Avg.\ precision \\
\midrule
AIBF bias magnitude $|\Delta|$   & \textbf{0.963} & \textbf{0.749} \\
Group membership                 & 0.672 & 0.118 \\
Ratio bias score (naive)         & 0.265 & 0.054 \\
\bottomrule
\end{tabular}
\end{table}

\subsection{Targeted review efficiency}
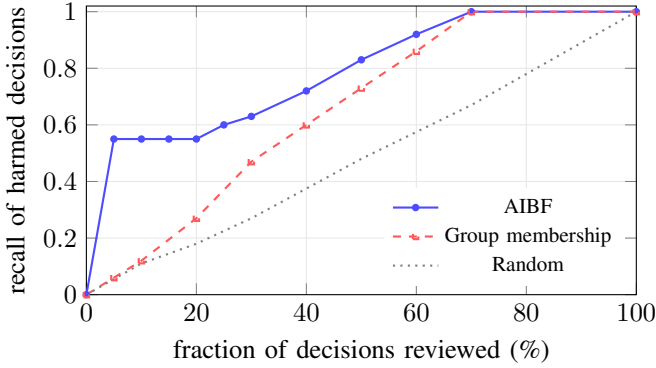
\begin{figure}[t]
\centering
\begin{tikzpicture}
\begin{axis}[
  width=\columnwidth, height=5.4cm,
  xlabel={fraction of decisions reviewed (\%)},
  ylabel={recall of harmed decisions},
  ylabel near ticks, xmin=0, xmax=100, ymin=0, ymax=1.02,
  legend style={at={(0.98,0.03)}, anchor=south east, font=\footnotesize, draw=none},
  grid=both, grid style={gray!18},
]
\addplot[blue!70, thick, mark=*, mark size=1pt] coordinates {
  (0,0)(5,0.55)(10,0.55)(15,0.55)(20,0.55)(25,0.60)(30,0.63)(40,0.72)(50,0.83)(60,0.92)(70,1.0)(100,1.0)};
\addplot[red!65, thick, dashed, mark=square, mark size=1pt] coordinates {
  (0,0)(5,0.06)(10,0.12)(20,0.27)(30,0.47)(40,0.60)(50,0.73)(60,0.86)(70,1.0)(100,1.0)};
\addplot[gray, dotted, thick] coordinates {
  (0,0)(10,0.11)(20,0.18)(30,0.27)(50,0.48)(70,0.67)(100,1.0)};
\legend{AIBF, Group membership, Random}
\end{axis}
\end{tikzpicture}
\caption{Recall of harmed decisions versus review budget on Adult. AIBF directed
review recovers most harmed candidates at a small budget.}
\label{fig:review}
\end{figure}

The practical value of localization is that a limited human review budget can be
directed to the decisions that most need it. Figure \ref{fig:review} plots the recall
of harmed decisions against the fraction of decisions reviewed, where review order is
set by AIBF bias magnitude, by group membership, or at random. Reviewing the five
percent of decisions with the largest AIBF magnitude surfaces fifty-five percent of
all harmed candidates, against six percent under group based review and six percent at
random. Table \ref{tab:review} gives the review budget needed to reach given recall
levels. To surface half of the harmed candidates an operator reviews one percent of
decisions under AIBF against thirty-two percent under group based review, a
thirtyfold reduction in review effort for the same protective coverage. The practical
significance is that human review is the binding constraint in a real audit, since a
compliance team can examine only a small fraction of decisions, and a method that
concentrates the truly biased cases at the top of the queue converts an obligation that
would otherwise be met by sampling into one that can be met by triage. The shape of the
AIBF curve, which rises steeply and then plateaus, is the shape a triage tool
should have, because it places nearly all of the harmed decisions in the first portion
of the queue and leaves a long tail of unaffected decisions that need no review.

\begin{table}[t]
\caption{Review budget (percent of decisions) needed to reach a given recall of
harmed decisions on Adult.}
\label{tab:review}
\centering
\footnotesize
\begin{tabular}{@{}lrr@{}}
\toprule
Target recall & AIBF & Group membership \\
\midrule
50\% & \textbf{1\%}  & 32\% \\
80\% & \textbf{34\%} & 57\% \\
90\% & \textbf{54\%} & 64\% \\
\bottomrule
\end{tabular}
\end{table}

\begin{figure}[t]
\centering
\begin{tikzpicture}
\begin{axis}[
  ybar, width=\columnwidth, height=4.7cm, bar width=6pt,
  ymin=0, ymax=44, ylabel={\% of decisions}, ylabel near ticks,
  symbolic x coords={0-2,2-5,5-10,10-15,15-20,20-30,30+},
  xtick=data, x tick label style={font=\scriptsize},
  enlarge x limits=0.09,
  legend style={at={(0.5,1.16)}, anchor=north, legend columns=2, font=\footnotesize, draw=none},
]
\addplot[fill=blue!30] coordinates {(0-2,41.09)(2-5,10.73)(5-10,12.67)(10-15,8.52)(15-20,10.48)(20-30,8.34)(30+,0.13)};
\addplot[fill=red!55] coordinates {(0-2,0.0)(2-5,0.16)(5-10,0.0)(10-15,0.51)(15-20,0.24)(20-30,7.09)(30+,0.04)};
\legend{not flagged, flagged}
\end{axis}
\end{tikzpicture}
\caption{Distribution of the AIBF bias magnitude $|\Delta|$ on Adult, in score points.
Flagged decisions concentrate at high magnitude, while the mass of unflagged decisions
sits near zero, so a magnitude threshold cleanly separates them.}
\label{fig:hist}
\end{figure}
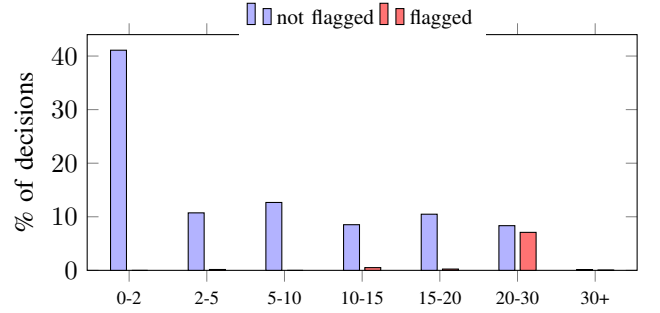

\subsection{Mitigation ceiling and proxy leakage}
It is tempting to close the loop by automatically correcting every flagged decision to
its counterfactual outcome. We measured the effect and report it in full, because it
reveals a limitation that operators must understand. On Adult, correcting all flagged
decisions raises the disparate impact ratio for sex from $0.078$ to $0.449$ and for
race from $0.386$ to $0.582$, a large improvement that nonetheless falls short of the
four-fifths threshold of $0.80$. On COMPAS the ratio rises from $0.645$ to $0.703$.
The reason the correction does not reach parity is proxy leakage: features labeled as
merit, such as education and hours worked on Adult, are themselves correlated with the
protected attributes, so neutralizing only the explicit protected proxies leaves a
residual disparity carried by the merit features. This is a property of the world and
the data, not a defect of the audit, and it is precisely why AIBF is positioned as a
detection and explanation tool that routes decisions to human judgment rather than as
an automated corrector. The finding also cautions against any method, including
group-blind approaches, that assumes removing explicit protected attributes yields a
fair decision.

\subsection{A worked audit}
Table \ref{tab:example} shows the audit of a real harmed candidate from the Adult test
set. The applicant scored $28.1$ and was rejected, but the counterfactual score, with
the protected proxies neutralized, is $66.3$, an advance, so the decision is a
harmed flip with a shift of $-38.1$ points. The explanation attributes the shift to
three protected factors, the largest being the female proxy at $-9.83$ points followed
by the age proxy at $-7.16$ and the minority race proxy at $-4.52$, while the merit
features contribute $+33.64$ points in the candidate's favor. An auditor reading this
does not receive a bare score but an actionable account: this qualified candidate,
whose merit features alone would advance them, was rejected because protected proxies
subtracted more than twenty one net points, and the decision warrants review. This is
the artifact a per-decision audit obligation calls for and that a group metric cannot
produce.

\begin{table}[t]
\caption{AIBF audit of a real harmed candidate (Adult test set). The protected proxies
subtract enough to flip an otherwise favorable decision.}
\label{tab:example}
\centering
\footnotesize
\begin{tabular}{@{}lr@{}}
\toprule
Component & Contribution (score pts) \\
\midrule
Merit features (education, hours, work class) & $+33.64$ \\
Protected: female proxy       & $-9.83$ \\
Protected: age-over-40 proxy  & $-7.16$ \\
Protected: minority-race proxy & $-4.52$ \\
\midrule
ATS score (reject)            & $28.1$ \\
Counterfactual score (advance) & $66.3$ \\
\textbf{Counterfactual shift} $\Delta$ & $\mathbf{-38.1}$ \\
\bottomrule
\end{tabular}
\end{table}

\section{Discussion}
\label{sec:disc}
The results are best understood not as a claim that AIBF is a better fairness metric,
which it is not and does not try to be, but as a claim that it occupies a position no
existing tool occupies, between the population level metric and the generic local
explainer, and that this position is the one that current audit obligations
require. The evaluation supports a specific claim about where AIBF fits. Against group fairness
metrics, AIBF adds localization and explanation: it identifies the individual
decisions behind an aggregate disparity and names their causes, which the metrics
cannot do and which per-decision audit mandates require. Against local explainers,
AIBF adds a protected versus merit partition and a counterfactual decision
level signal that is tied to whether the outcome actually changed, which raw
attribution magnitude is not, as the collapse of the ratio baseline to below chance
demonstrates. The counterfactual construction is what unifies the two levels, because
its per-decision values are faithful to the group disparity while remaining
individually actionable.

The review efficiency result is the most consequential for practice. Audit mandates
create a real operational problem, since a human cannot review every decision, and an
operator who reviews by group membership both wastes effort on unaffected candidates
and, as our correction experiment shows, risks overcorrecting. AIBF turns a population
level obligation into a ranked, explained worklist, which is the form in which a
compliance team can act on.

\subsection{Deployment}
AIBF is intended to run in one of two modes. In batch audit mode it processes a
period's decisions, produces the disparate impact report required by regulation, and
attaches the ranked worklist of flagged decisions with their explanations, which is
the artifact an independent auditor under Local Law 144 needs. In inline mode
it audits each decision as it is made and routes a flagged decision to human review
before a rejection is finalized, which is the posture the human oversight requirement
of the EU AI Act encourages. Both modes use the same two queries per decision and the
same explanation machinery, and both keep the protected signals strictly on the audit
side of the system, never returning them to the scorer.

\subsection{Contestability}
A per-decision, explained audit does more than direct review. It makes a decision
contestable, which is the affordance that both the counterfactual explanation
literature \cite{wachter} and data protection law \cite{gdpr} treat as central to
automated decision making about people. A candidate told only that an opaque system
rejected them has nothing to challenge. A candidate whose rejection carries a
counterfactual, that the decision would have been an advance had protected proxies not
lowered the score by a stated number of points, together with the factors responsible,
has a specific and answerable account. AIBF produces that account as a byproduct of the
same computation that drives detection, at no additional cost.

\subsection{A lesson for fairness attribution}
The collapse of the ratio baseline to below chance points to a general
lesson, because a ratio of protected to total attribution magnitude is an intuitive and
tempting definition of per-decision bias, and it is the definition an earlier version of
this work used. The finding is that any per-decision fairness signal computed from
attribution magnitudes alone, without reference to the decision boundary, can rank the
decisions that matter in the wrong order, because influence that does not move a
decision across the threshold is not the influence an audit cares about. The
counterfactual construction is the natural fix, since it measures influence in the units
of the decision itself. We suspect this observation applies beyond our particular score,
to any attempt to summarize local fairness attributions into an actionable per-decision
quantity, and we offer it as guidance to others building such tools.

\subsection{Recommendations for auditors}
The findings translate into concrete guidance for anyone auditing an automated hiring
system, whether or not they adopt this implementation.

\begin{itemize}
\item \textbf{Audit at the decision level, then aggregate.} Compute the per-decision
counterfactual first and derive the group report from it, rather than computing only the
group metric, so that the aggregate always comes with the localized evidence behind it.
\item \textbf{Define bias relative to the decision boundary.} Do not rank decisions by
raw or ratio attribution magnitude, which ignores the margin and can invert the cases
that matter; rank by the counterfactual shift, which is in the units of the decision.
\item \textbf{Prefer targeted review to blanket review.} Reviewing by group membership
wastes scarce human attention on unaffected candidates and risks overcorrection; a
ranked worklist directs review to the decisions protected attributes changed.
\item \textbf{Treat the correction ceiling as a warning about proxies.} If neutralizing
explicit protected attributes does not reach parity, the residue is proxy leakage in the
merit features, which should prompt scrutiny of those features rather than confidence in
a group-blind design.
\item \textbf{Calibrate the threshold locally and keep protected data on the audit
side.} Tune the flag threshold on a labeled sample for the specific deployment, and
ensure the protected attributes used for auditing never re-enter the scorer.
\end{itemize}

\subsection{Generality beyond hiring}
Although we motivate AIBF by automated hiring, nothing in the method is specific to it.
The audit requires only a thresholded scoring decision, a partition of features into
protected proxies and legitimate factors, and query access to the scorer, which are
present in lending, insurance underwriting, admissions, tenant screening, and pretrial
risk assessment. The COMPAS result is direct evidence of this generality, since it is a
criminal justice scorer rather than a hiring one, and AIBF audits it with the same
faithfulness, the same clean separation of flipped decisions, and the same
correction ceiling. What transfers is the core observation that a per-decision fairness
signal should be defined by a counterfactual on the protected partition relative to the
decision boundary, which turns a group level obligation into an individually
actionable, explainable audit in any of these domains. We report hiring results because
that is where the per-decision audit mandate is most explicit today, but the method is
a general instrument for auditing thresholded decisions about people.

\section{Limitations and Threats to Validity}
\label{sec:limits}
We state the limits directly, since several bound the strength of the results.

\begin{itemize}
\item \textbf{Model based, not causal.} AIBF intervenes on the deployed scorer's
inputs rather than on a structural causal model of the world \cite{kusner}, so it
measures the influence the scorer places on protected proxies. This is the right
target for auditing a specific deployed system, but it is not a claim about real world
causation, and a proxy that the scorer reads only indirectly is captured only to the
extent the neutralized features carry it.
\item \textbf{Proxy coverage.} AIBF audits the proxies it is given. Proxies it does
not model, and the residual proxy leakage documented in Section \ref{sec:eval}, limit
both detection and any correction, which is why the corrected disparate impact ratio
does not reach parity.
\item \textbf{Linear reference for explanations.} The itemized explanation uses a
linear reference model, which makes the additive attribution exact for that reference
but under-describes a strongly non-linear scorer. The decision signal $\Delta$ is
unaffected, since it is computed on the true scorer.
\item \textbf{Audited systems are our own.} We audit standard classifiers trained on
real data rather than proprietary production systems, which we cannot access. The
disparate impact these systems exhibit is real, but the magnitudes on a given vendor
system would differ, and only a study with scoring access could measure them.
\item \textbf{A signal, not a verdict.} A high bias magnitude is a trigger for human
review, not a legal determination of discrimination, and the paper's numbers should
not be read as adjudicating any specific system.
\end{itemize}

\subsection{Threats to validity}
We separate three questions because they bear differently on the claims. The first,
construct validity, is whether the counterfactual flip is the right target. It is the
operational analogue of counterfactual fairness for a deployed scorer \cite{kusner},
and it gives an auditor what they need, a flagged decision that arrives with a concrete
witness of the change, though it captures influence within the model and not real world
causation. The second, internal validity, is whether the comparison is fair, and here
the AIBF signal, the group baseline, the margin baseline, and the ratio are all scored
against the same flips on the same held out data with the same scorer, so the gaps
between them reflect the signal and not the setup. The third, external validity, is
where the claims are weakest: we trained the audited scorers ourselves, the datasets
number two, and one lies outside hiring, so the specific magnitudes will not carry over
to a particular production system. What does carry over is the shape of the findings,
that the counterfactual signal is faithful, that it beats the group baseline at
localizing biased decisions, and that the ratio fails, since these follow from the
construction rather than from the data at hand.

\subsection{Reproducibility}
Every number in this paper is produced by the released scripts from public datasets
with a fixed random seed. The data preparation, the audited classifier, the
counterfactual audit, the detection and review-efficiency analyses, and the figure
data are each a single command, so the results can be regenerated and inspected rather
than taken on trust. We regard this as essential for an auditing method, whose own
claims should be as checkable as the decisions it audits.

\section{Ethical and Legal Considerations}
\label{sec:ethics}
AIBF is designed to support human and legal judgment rather than to replace it. Its
per-decision, counterfactual reasoning is compatible with the selection rate and
impact ratio reporting required by New York City Local Law 144 \cite{ll144} and with
the four-fifths rule \cite{uniform}, and its transparency and human oversight posture
align with the EU AI Act's obligations for high risk recruitment AI \cite{euaiact}
and with the contestability aims behind counterfactual explanation \cite{wachter} and
data protection law \cite{gdpr}. Because the protected signals AIBF uses are
themselves sensitive, the method treats them strictly as audit inputs and never as
scoring inputs, and it recommends that flagged decisions route to human review rather
than to automatic action, since a detector that misfires would otherwise harm the very
candidates it is meant to protect. All experiments in this paper use public research
datasets and contain no private candidate data.

The audit itself must be used responsibly. A bias magnitude threshold is an operating
point, and setting it too aggressively floods human reviewers and risks second guessing
sound decisions, while setting it too permissively misses harm; the threshold should be
calibrated on a labeled sample for each deployment rather than carried over from another
setting, a caution reinforced by our own finding that a threshold tuned on one
distribution transferred poorly. Because AIBF consumes protected attributes to perform
its audit, those attributes must be handled under the same protections as any sensitive
data and must never be allowed to flow back into the scoring path, which the design
enforces by keeping them strictly on the audit side. Finally, an audit that improves the
efficiency of review is not a substitute for the human judgment it informs, and its
outputs are evidence for a person, not a verdict.

\section{Conclusion and Future Work}
\label{sec:conc}
We presented AIBF, a counterfactual per-decision method for auditing bias in automated
hiring that produces a signed bias in score points, a flag for the decisions protected
attributes changed, and an explanation of why. On two real datasets the per-decision
signal is faithful to the group disparity, detects biased decisions far better than a
group membership baseline, and turns a population level audit obligation into an
efficient, explained review worklist. We also showed that attribution based
correction improves the disparate impact ratio without reaching parity, because merit
features carry residual proxy correlation.

Future work proceeds along four lines. The first is a causal formulation that
intervenes on a structural model rather than on the scorer's inputs, tightening the
counterfactual. The second is proxy discovery, learning which nominally neutral
features leak protected information so that the audit and any correction can account
for them. The third is a study on live proprietary applicant tracking systems through
their scoring interfaces, to measure disparate impact and audit efficiency in
production. The fourth is a human study of whether the explanations change reviewer
decisions in practice. The implementation, the datasets' preparation, and the scripts
that produce every number in this paper are available under the Apache 2.0 license at
the address on the title page.

\end{document}